\documentclass{svjour2}                    % onecolumn
\smartqed  % flush right qed marks, e.g. at end of proof
\usepackage{graphicx}
\newcommand{\no}{\nonumber}
\def\lsim{\mathrel{\rlap{\lower4pt\hbox{\hskip1pt$\sim$}}
    \raise1pt\hbox{$<$}}}         %less than or approx. symbol
\def\gsim{\mathrel{\rlap{\lower4pt\hbox{\hskip1pt$\sim$}}
    \raise1pt\hbox{$>$}}}         %greater than or approx. symbol

\newcommand{\sgn}{\mathop{\rm signum}}
\journalname{}
\begin{document}

\title{DNA Twist Elasticity: Mechanics and Thermal 
Fluctuations%\thanks{Grants or other notes
}


\author{Supurna Sinha         \and
        Joseph Samuel %etc.
}


\institute{Supurna Sinha\at
              Raman Research Institute, Bangalore, India\\
              Tel.: +91-80-23610122\\
              Fax: +91-80-2361 0492\\
              \email{supurna@rri.res.in}           %  \\
           \and
           Joseph Samuel\at
            Raman Research Institute, Bangalore, India
} 

\maketitle

\begin{abstract}

The elastic properties of semiflexible polymers are of great
importance in biology. There are experiments
on biopolymers like double stranded DNA, which twist and stretch single
molecules
to probe their elastic properties. It is known that thermal fluctuations
play an important role in determining molecular elastic properties, but
a
full theoretical treatment of the problem of twist elasticity of
fluctuating ribbons using the simplest worm like chain model
(WLC) remains elusive.
In this paper, we approach this problem by taking first a mechanical
approach and then incorporating thermal effects in a quadratic
approximation applying the Gelfand-Yaglom (GY) method for computing
fluctuation determinants. Our study interpolates between mechanics
and statistical mechanics in a controlled way and shows
how profoundly thermal fluctuations affect  the elasticity of
semiflexible polymers.
The new results contained here are:
1) a detailed study of the minimum energy configurations with
explicit expressions for their energy and writhe and plots of
the extension versus Link for these configurations.
2) a study of fluctuations around the local minima of energy and
approximate analytical formulae for the free energy of
stretched  twisted polymers derived by the Gelfand Yaglom method.
We use insights derived from our mechanical approach to suggest
calculational schemes that lead to an improved treatment of thermal
fluctuations.
From the derived formulae, predictions of the WLC model
for molecular elasticity can be worked out for comparison against
numerical simulations and experiments.

\end{abstract}

\section{Introduction}
\label{intro}

In recent years, the theoretical study of the 
elasticity of semiflexible polymers has emerged as an active area of 
research. These studies are motivated by 
micromanipulation experiments\cite{bust,sethna,sethnawang,sethnadaniels} 
on biopolymers in which single molecules are
stretched and twisted to measure elastic properties. These
experiments are designed to understand the role of semiflexible polymer
elasticity in the packaging of these polymers in a cell
nucleus. Twist elasticity plays an
important role in several biological functions. DNA, a long 
molecule (between microns and meters in length) which carries 
the genetic code, is neatly packed into the tiny cell nucleus, just a 
few microns across.  
The first step in packaging DNA in a cell nucleus 
involves DNA-histone association which makes use of supercoiling 
in an essential way. The process of DNA
transcription can generate and be regulated by supercoiling\cite{strick}.

The single molecule experiments have been theoretically analyzed 
using the popular Worm Like Chain model, which gives an excellent 
account of the bending elastic properties of DNA \cite{siggia}. When the
twist degree of freedom is explored by rotationally constraining the 
molecule the theoretical analysis gets complicated by subtle topological and 
differential geometric issues.
In a previous work \cite{samsupabhi} we have treated these 
subtleties in the  language
of differential geometry and topology. 
Our present study builds on earlier 
mechanics based research in this 
area\cite{fain,maddox,fainrudnickostlund,sethnadaniels} and goes on to 
incorporate the effects of thermal fluctuations in a quadratic 
approximation. We evaluate the partition function and 
present an explicit analytical form which can be compared against
measured elastic properties or numerical simulations 
\cite{neukirchmech}. 

Two parallel streams of research have evolved in the study of semiflexible
polymer elasticity. Some researchers use classical elasticity which is
conceptually simple, but ignores thermal fluctuations which are important 
to the problem. Others use a statistical mechanical approach which 
properly takes into account thermal fluctuations and entropic effects.
This latter approach is beset by the geometrical and topological 
subtleties
we alluded to earlier. In the real biological context of a cellular
environment, these polymers
get constantly jiggled around by thermal fluctuations and therefore a
statistical mechanical approach is necessary. Biological processes involve
the entire range of flexibilities: Actin filaments, microtubules
and short DNA strands are energy dominated, while long DNA strands
have appreciable entropy. 
The quantitative measure of flexibility is
$L/L_P$, where $L$ is the contour length of the polymer and $L_P$ its 
persistence
length ($16\mu$ for Actin and $50 \rm nm$ for DNA). In the
energy dominated short polymer regime classical elasticity gives a fair
understanding of the problem. In the absence of external forces 
and torques, long polymers require a statistical mechanical treatment 
since the entropic contribution is not small. 
In this paper we consider 
polymers subject to a stretching force and torque and develop a theory 
of 
thermal fluctuations around the minimum energy configurations.
These minimum energy configurations come in two families, the straight
line family (SLF) and the writhing family (WF). 
We present a treatment based on classical elasticity and 
then incorporate thermal fluctuations around the minimum energy 
configurations. Members of the writhing family 
are far from straight and so our analysis goes far beyond perturbation 
theory about the straight line. While our general analysis applies to 
polymers of all lengths, some of our analytical formulae are derived in 
the long polymer limit. Our treatment is analytical and 
complementary to numerical studies\cite{volo,volosimu,neukirchmech}. 

Two extreme situations are relatively well understood. If the polymer is nearly 
straight, one can use perturbation theory \cite{nelson,writhe,abhijit} 
about the 
straight line to calculate its elastic properties. 
In the opposite extreme when the 
polymer is wrung hard, it buckles and forms plectonemic structures 
\cite{marko,fuller,fuller1} which are stabilized by the finite thickness 
of the DNA. 
The transitional regime where the polymer is neither straight nor
plectonemic is not as 
easy. Perturbation theory about the straight line is not applicable and 
neither do we benefit from simplifications that arise from the energy 
dominated plectonemic regime. 
More specifically, let the stretching force $F$ be 
$2k_BT/L_P$ or higher. As we turn the bead to twist or ``wind up'' the 
molecule, it initially remains approximately straight. At high links
the polymer buckles and winds around itself to form plectonemic 
structures.
The intermediate link regime, in which the polymer is
neither straight nor plectonemic is the subject of this paper. 
(These three regimes of low link, intermediate link and high link
have also been referred to in \cite{neukirchmech} as
straight line, buckling and supercoiled regimes.)
We will study the straight line and the
writhing solutions and fluctuations around these solutions.

The central goal of this paper is to derive explicit analytic 
expressions 
for the free energy and the writhe of a twisted stretched 
polymer. Derivatives 
of the free energy expressions computed here connect to 
measurable quantities which can be probed via single molecule experiments
or simulations. 
We arrive at the free energies by focusing on the stationary points of 
the energy functional dominating the partition function. 
While there is interest in the elastic properties of DNA at all length 
scales, the experiments \cite{bust,strick,sethnawang}
deal with long polymers, many times
the persistence length of about $50 \rm{nm}$. 
Many theoretical treatments 
\cite{neukirch,Bouchiat,maggs,neukirchmech}
also focus on this regime of long polymers, which leads to 
some simplification.

In order to formulate the problem we start from an idealized experiment. Let 
us imagine that a twist storing semiflexible polymer (like dsDNA or 
filamentary Actin) is attached to a
glass slide at one end and a bead at the other, so that its tangent vector 
at  both ends is constrained to point in the $\hat{z}$ direction.
Supposing
the glass slide end to be at the origin, we apply a force $F$ and torque 
$\tau$ (both in the $\hat{z}$ direction) on the bead. The elastic response of 
the polymer can be described by its extension $z=\vec{r}.{\hat F}$ (the 
position 
of the bead) and its link (the number of times the bead has turned about 
the $\hat{z}$ axis).
Real experiments on DNA differ from the one described above mainly 
in that they deal with long polymers and it is usually 
the Link rather than the torque that is held constant. 
Though the experiments \cite{sethnawang,laporta} could easily be
adapted by using a feedback loop to maintain constant torque.

Actually, we will be interested in a wider class of thought experiments 
of which the above is a particular realization. We may, for instance, 
use more general boundary conditions, fixing the tangent vector at the ends 
to ${\hat t}(0)={\hat t}_i$ and ${\hat t}(L)={\hat t}_f$. 
(The force is still along ${\hat z}$ and the torque along ${\hat t}_f$.)
As theorists, we may also explore the entire range of 
flexibilities  and parameters, which are not practical in the currently 
accessible regime. Our study is not restricted to long polymers, though
we will sometimes specialize to this case for simplicity. We will also
regard all our parameters (contour length, persistence length, force, 
torque,
temperature, elastic constants and boundary conditions) as tunable 
without worrying about how this may be experimentally achieved. In the 
mechanical limit our analysis reduces to the classical theory of beams
and cables. Our larger aim is a 
comprehensive understanding of the entire range of
parameters covering the range from beams and cables to DNA. 
Thus we are not immediately concerned with 
any one experiment, but a class of them. 
Detailed treatments of actual experiments are contained in Refs. 
\cite{marko,sethnadaniels}. For example these treatments take into 
account the chiral nature of the DNA molecule, which we ignore. We are 
more
concerned here with mathematically working out the predictions of the 
simplest worm like chain model rather than a detailed 
modelling of
specific experimental data. However, our focus is  
experimental relevance and we calculate from the model, 
quantities like 
link-extension relations which are measured in laboratories. 

The paper is organized as follows. 
Sec $II$ deals with the mechanics of semiflexible polymers.
We start by noting that the problem of computing the partition 
function of a twist storing polymer separates into two parts, a simple 
Gaussian over the twist and a harder problem involving the writhe.
We derive the Euler-Lagrange (E-L) equations describing the minimum 
energy configurations. The solutions of the E-L equations 
can be expressed using elliptic integrals. 
Sec $III$ deals with the role of thermal fluctuations.
We perform the second variation of the energy and 
explain the methods we use to compute the thermal correction to the 
free energy due to fluctuations about the 
classical solutions. 
This gives the main result of our paper, an approximate 
expression for the free energy of a stretched twisted polymer 
which takes into account thermal fluctuations around nonperturbative
solutions.  We then suggest 
calculational schemes (Sec. $IV$  ) for 
a more ambitious treatment of thermal fluctuations.
We end with 
some concluding remarks in Sec. $V$.  
\section{Mechanics}
\label{mechanics}
We model the twist storing polymer by a ribbon $(\vec{x}(s),\; 
{\hat e}^{i}(s))$, 
($i=1,2,3$) which is a framed\footnote{Note that we {\it do not} use 
Serret-Frenet framing, which is popular in this field. Serret-Frenet
framing becomes ill defined over 
locally straight 
pieces of the curve.} space curve. $\vec{x}(s)$ 
describes 
the 
curve, $\hat{t}(s)  =  \frac{d\vec{x}}{ds}$, its tangent vector 
and $e^{i}(s)$ the framing. $s$ is the arc length parameter along the 
curve ranging from $0$ to $L$ the contour length of the curve. 
${\vec 
x}(0)=0$ since one end 
is 
fixed at the origin. The tangent 
vectors at both ends ${\hat t}(0)$,${\hat t}(L)$ are fixed. 
A  force ${\vec F}=F{\hat z}$ along the ${\hat z}$ direction is applied 
at the free end at ${\vec x}(L)$ and also a torque $\tau$ along the 
fixed end tangent vector ${\hat t}(L)$.
We suppose $\hat{e}_{3} = \hat{t}(s)$ and refer to 
$\hat{e}_{1}(s)$ 
as the ribbon vector that describes the twisting of the polymer about its 
backbone. Defining the ``angular velocity'' $\vec{\Omega}$, via
\begin{equation}
\frac{d\hat{e}_{i}}{ds} = \vec{\Omega} \times \hat{e}_{i}
\label{ang}
\end{equation}
The ``body-fixed angular velocity'' components of ${\vec \Omega}$ are
\(\Omega_{i} = \hat{e}_{i} \cdot \vec{\Omega}\). 
The expression for the energy of a 
configuration is 
\begin{equation}
{\cal E}_0 ({\cal C}) = \int^{L}_{0} \frac{1}{2} [A(\Omega_{1}^{2} + 
\Omega^{2}_{2}) + C\Omega^{2}_{3}] ds - \int^{L}_{0} \vec{F}.\vec{t} ds 
- 2\pi \tau {\rm Lk}, 
\label{energyfull}
\end{equation}
where ${\rm Lk}$ is the number of times the bead is turned around the 
${\hat t}_f$ axis.
The mathematical problem we face is to compute the partition function
\begin{equation}
Z_0(F, \tau ) = \sum_{{\cal C}} \exp - \frac{{\cal E}_0({\cal 
C})}{k_B T}\;. 
\label{ptnfull}
\end{equation}
In Eq.(\ref{ptnfull}), the sum is over all allowed configurations of the 
ribbons \cite{samsupabhi}. The ribbon can be closed with a fixed 
reference ribbon that goes a long way 
in the $\hat{t}_f$ direction, makes a wide circuit and 
returns to the 
origin along the $\hat{t}_i$ direction\footnote{
The reference ribbon is supposed to be nowhere self intersecting 
or south pointing \cite{samsupabhi}. So we also suppose 
${\hat t}_i, {\hat t}_f\ne -{\hat z}$.}.
In the 
calculations below, we will 
set both $A$ and $k_BT$ equal to unity and restore them  
when necessary.

We use the celebrated relation\cite{white,calugeranu} decomposing the link into 
twist and writhe:
\begin{equation}
{\rm Lk = Tw} + {\cal W}_{CW}   
\label{celeb}
\end{equation}
where ${\rm Tw} = 2\pi\int{{\Omega_3(s)}ds}$
and ${\cal W}_{CW}$ is the writhe.
The writhe is a non-local quantity defined on closed simple
curves: Let the arc length parameter $s$ range over the entire length
$L_0$ of the closed ribbon
(real ribbon $+$ reference ribbon) and
let us consider the curve ${\vec x}(s)$ to be a periodic function
of $s$ with period $L_0$.
Let ${\vec R}(s,\sigma)={\vec x}(s+\sigma)-{\vec x}(s)$.
The C\u{a}lug\u{a}reanu-White
writhe is given by\cite{dennis,calugeranu,white,fuller1,fuller}
\begin{equation}
{\cal W}_{CW}= \frac{1} 
{4\pi}\oint_{0}^{L_{0}}ds\int_{0+}^{L_{0}-}d\sigma
[\frac{d{\hat{R} (s,\sigma)}}{ds}
\times \frac{d{\hat{R}(s,\sigma)}}{d\sigma}]\cdot{\hat{R}.}
 \label{writhe}
 \end{equation}
Because of White's theorem Eq.(\ref{celeb}), we find that the problem 
neatly splits 
\cite{fuller,nelson,samsupabhi} 
into two parts
\begin{equation}
Z_0(F, \tau) = Z_{1} (F, \tau) Z_{2}(\tau) 
\label{rep} 
\end{equation}
where
\begin{equation}
Z_{1} (F, \tau) = \int {\cal D}[\vec{x}(s)] \exp - {\cal 
E}_{1}[\vec{x}(s)] 
\label{part1}
\end{equation}
and
\begin{equation}
{\cal E}_{1}[\vec{x}(s)] = \int^{L}_{0} ds\; \frac{1}{2}\; 
\frac{d\hat{t}}{ds} \cdot \frac{d\hat{t}}{ds}\; - \int^{L}_{0} \vec{F} \cdot \hat{t} ds 
- 2\pi \tau {\cal W}_{CW} .
\end{equation}
$Z_{2}(\tau)$ is given by a Boltzmann sum over framings 
with 
${\cal E}_{2}$ given by
\begin{equation}
{\cal E}_{2}(\tau) = \int ds [\frac{C}{2} \Omega^{2}_{3} - \tau 2\pi 
{\rm Tw} 
]
\end{equation}
$Z_{2}(\tau)$ is easily evaluated as a Gaussian integral and gives 
\begin{equation}
Z_{2}(\tau) = \exp{[\frac{\tau^{2}L}{2C}]}=\exp{[-{\cal 
G}_2(\tau)]}.
\label{twistenergee}
\end{equation}
Thus, the problem that remains is to compute 
$Z_{1} (F, \tau)$ (Eq.(\ref{part1}) )
which depends only on the curve $\vec{x}(s)$ and not its framing. 

This problem is hard because of the appearance of the writhe 
${\cal W_{CW}}[\vec{x}(s)]$, which
is a non local function of the curve. 
However, we will make progress by noting that
{\it variations} of the writhe are local\cite{samsupabhi,papa}.
We will compute the partition function assuming that for high 
stretch forces, 
the sum over curves is dominated by configurations near minima of the energy. 
The 
approximation consists of using an expansion of the energy about the minimum 
energy configuration and keeping fluctuation terms about the minimum 
to quadratic order.
This section deals with the minimum of energy and the following one with 
fluctuations. We thus derive explicit analytical expressions for the 
Gibbs free energy of the polymer ${\cal G}(F, \tau)$ which can be used 
to 
compute 
its elastic response. 

Theoretically, it is easiest to 
deal with the constant torque ensemble. In the quadratic approximation, 
the conjugate ensembles are equivalent 
\cite{keller,ineq,bionel,freesinha}
even for polymers of finite length. This equivalence of ensembles holds 
exactly for long polymers, since these are at the thermodynamic limit.
(See however, the remarks below concerning stability in different 
ensembles.)

In this section, we ignore thermal fluctuations and 
consider 
a purely classical mechanical analysis
to study the configurations
of a torsionally constrained 
stretched semiflexible biopolymer like DNA. 
While such treatments exist in the literature 
\cite{fuller,rabin,love,maddox,fain,fainrudnickostlund}, we will present 
a slightly 
different perspective based on analogies with classical mechanics,
which help us to incorporate
thermal fluctuations. 

The central quantity of interest is the bending, stretching and 
writhing energy,
\begin{equation}
 {\cal E}({\cal C}) = \frac{1}{2} \int_0^L \dot{\hat t}.\dot{\hat t} ds - 
\int_{0}^{L} \vec{F}.\hat{t} ds - 2\pi \tau {\cal W}_{CW} 
\label{energy}
\end{equation}
of a space curve $\vec{x}(s)$ whose tangent vector is 
$\hat{t}(s) = \frac{d\vec{x}}{ds}$. 
The tangent vector is varied subject to the boundary conditions 
$\hat{t}(0) = \hat{t}_i$,
$\hat{t}(L) = \hat{t}_f$ 
fixing the tangent vector at both ends of the curve. Remembering that
the {\it variations} of writhe are local\cite{samsupabhi,papa}, we 
arrive at
the Euler-Lagrange equations (see Eq.(\ref{ELprime}) below) from 
Eq.(\ref{energy}): 
\begin{equation}
-\ddot{\hat t} - 
{\vec F}=\tau({\hat t}\times{\dot{\hat t}})- \gamma{\hat t},
\label{eqnofmot}
\end{equation}
where the term $\gamma{\hat t}$ arises 
since ${\hat t}\cdot{\hat t} = 1$. 
The problem is formally similar to a symmetric top, a fact that was well 
known to 
Kirchoff
%\cite{kirchoff}. 
The analogy is useful for integrating the 
Euler Lagrange 
equations. We use quotes for the analogous top quantities. The ``Kinetic 
energy'' is given by 
$T=\frac{1}{2}{\dot{\hat t}} \cdot \dot{\hat t}$
and the ``potential energy'' is
$V={\vec F}\cdot {\hat t}.$
The total ``energy''
\begin{equation}
{\cal H} = T + V = \frac{1}{2} \dot{\hat t}\cdot\dot{\hat t} + 
\vec{F}\cdot{\hat t}=\frac{1}{2} (\dot{\theta}^2+\sin^{2}\theta\dot{\varphi}^2)+F\cos\theta
\label{hamex}
\end{equation}
is a constant of the motion as is the $z$ component of the ``angular momentum''
\begin{equation}
J_{z}=(t \times\dot{\hat t})_z-\tau{\hat t}_{z} = \sin^{2}\theta\dot{\varphi} 
- \tau\cos\theta,
\label{zang}
\end{equation}
where we have introduced the usual polar coordinates on the sphere of 
tangent vectors. Using these ``constants of the motion'' we reduce the 
problem to quadratures as described in \cite{goldstein}. The basic 
equations
are 
\begin{equation}
\frac{\dot\theta^{2}}{2} = {\cal H} - F \cos\theta - 
\frac{(J_{z} + \tau\cos\theta)^{2}}{2\sin^{2}\theta} 
\label{hamex1}
\end{equation}
\begin{equation}
\dot{\varphi} = \frac{(J_{z} + \tau\cos\theta)}{\sin^{2}\theta}.
\label{phidot}
\end{equation}
Setting $u = \cos\theta$, we find that 
\begin{equation}
{\dot u}= \epsilon f(u) 
\label{solu}
\end{equation}
where $f(u)=\sqrt{{\cal P}(u)}$,$\epsilon=\pm 1$  and 
\begin{equation}
{\cal P}(u):=2({\cal H}-Fu)(1-u^2)-(J_z+\tau u)^2
\label{polygeneral}
\end{equation}
is a cubic polynomial in $u$. Integration of Eq.(\ref{solu}) gives $u(s)
$ in terms of elliptic integrals and Eq.(\ref{phidot})
gives $\varphi(s)$. Further integrating the tangent vector
gives us the polymer configuration ${\vec x}(s)$ in real space.
Since our polymer is under stretch (unlike that of Ref. \cite{emanuel}   
which is under compression) we assume $F$ is positive.
${\cal P}(u)$ is 
positive for large positive $u$, and 
negative (or zero) at $u=1$ and $u=-1$. 
In order for there to be a physical solution there must be 
two {\it real} roots, $b,c$ (turning points) in the physical range of 
$u=\cos{\theta}$: $-1\le c\le b \le1$ and  
a third real root $a\ge1$.
We have ordered the three roots so that
$a\ge b\ge c$. 
The boundary data must lie between $c$ and $b$. $c \ge u_i,u_f \le b$ . 
The ``motion'' goes from $u_i$ to $u_f$, possibly passing through 
turning points on its way.

The integration constants ${\cal H}$ and ${J_z}$ are determined by 
the length of the polymer and the boundary conditions.
From Hamilton-Jacobi theory  we can write the energy  of the solution of 
length $L$ going from 
$u_i,\varphi_i$ to $u_f,\varphi_f$ as
\begin{equation}
{\cal E}(u_i,u_f,\varphi_i,\varphi_f,L)=
\int_{u_i}^{u_f}du f(u) -{\cal H}L +J_z\Phi
\label{action}
\end{equation}
where $L$ and $\Phi=\varphi_f-\varphi_i$,
given by
\begin{equation}
L=
\int_{u_i}^{u_f}\frac{du}{f(u)} 
\label{length}
\end{equation}
and
\begin{equation}
\Phi=
\int_{u_i}^{u_f}\frac{du} {f(u)} h(u)
\label{phiaction}
\end{equation}
determine ${\cal H}$ and $J_z$ and 
$h(u)=(J_z+\tau u)/(1-u^2)$. 
These integrations must include turning points if any. For instance,
if $u$ goes from $u_i$ to $u_f$ via $c$, 
$\int_{u_i}^{u_f} = \int_{c}^{u_i}+\int_{c}^{u_f}$,
allowing for sign changes in the integrand Eq.(\ref{solu}).

We will use these relations
below in computing the fluctuation determinant in the next section. 
The expressions for 
$\Phi$, $L$ and ${\cal E}$ are elliptic integrals, but as 
these 
functions are not as widely known today as they were a century ago, 
it is more useful and appropriate to leave them as integrals. 
It is quite easy to numerically evaluate these 
integrals on a computer and plot any function which may be of interest.

\subsection{Symmetric Boundary Conditions:} 
\label{symmetric}
As an illustrative 
example, 
we specialize to boundary conditions ${\hat t}_i={\hat t}_f={\hat z}$.
These boundary conditions
imply that $\theta = 0$ is a point on the solution. Since 
${\dot\theta}^{2}$ has 
to be finite and non-negative at $\theta = 0$, we have $J_{z} = -\tau$
and ${\cal H}\ge F$.
The form of 
${\cal P}(u)$ simplifies to 
$${\cal P}(u)=(1-u)[2({\cal H} - {F} u)(1 + u) - \tau^{2} (1-u)],$$
and Eq.(\ref{phidot}) reduces to 
\begin{equation}
\dot{\varphi} = -\frac{\tau}{1+u}.
\end{equation}
We want to find the minimum energy configurations 
subject to the writhe constraint.
These configurations satisfy the Euler-Lagrange equations. The simplest 
solution is the straight line ${\hat t}(s) = {\hat z}$ or $\theta(s) = 0$
for all $s$. This configuration globally minimizes 
${\cal E}({\cal C})$ 
\begin{equation}
{\cal E_{ST}} =  - FL
\label{stsolu}
\end{equation} 
but since ${\cal W}_{CW} = 0$, it cannot 
accommodate writhe. For this we need the ``writhing solution'',
in which 
the tangent vector ${\hat t}$ starts from ${\hat z}$ at $s=0$ 
and $\theta$ increases to a maximum of $\theta_0$ at $s=s_0$, the 
turning point and then returns to ${\hat z}$ at $s=2s_0$. The period of
the orbit is $P=2s_0$. As was shown in \cite{papa}, the configuration 
is a local minimum of the energy only if the length L of the polymer
is equal to $P$ (and not a multiple of it).

It has been shown earlier \cite{papa} 
that 
the local minima of the 
energy are ``good curves'' \cite{neukirch} {\it i.e} they satisfy the 
conditions of Fuller's \cite{fuller,fuller1} 
theorem and so the writhe can be computed from a 
simple local formula \footnote{As a side remark we mention that our 
approach is guided by geometric
phase ideas\cite{shapere,maggsberry,js,twist}}.

\begin{equation}
{\cal W}_{CW}=
\int_{u_i}^{u_f}\frac{du(1-u)h(u)} {f(u)} 
\label{writheexp}
\end{equation}
Below we will sometimes 
drop the subscript on ${\cal W}_{CW}$, it is understood
that we only deal with ``good curves''.

\subsection{Long Polymers}
\label{longpolymers}
One gets further simplification by confining to the limit of long 
polymers. In this limit,  the ``period'' $P$ 
($=L\rightarrow\infty$). As a result the doubly periodic functions that
appear for finite $L$ become simple trigonometric functions (since one 
of the periods tends to infinity). We will briefly restrict our attention
to infinitely long polymers in this section to arrive at
an explicit analytical form for the ``writhing family'' of solitons.
To reach this limit, the 
length Eq.(\ref{length})
must diverge. 
For ${\cal H}>F$, the function ${\cal P}(u)=f^2(u)$ 
vanishes linearly 
as $u\rightarrow 1$. The corresponding integral converges and does not lead to  
$L\rightarrow\infty$. Thus we must require that ${\cal P}(u)$ vanishes 
quadratically as  $u \rightarrow 1$ so that 
$\int^{1}_{u_{0}}{\frac{du}{f(u)}}$
diverges logarithmically at the upper limit. This situation obtains
if ${\cal P}(u)$ has coincident roots at the upper limit. 
Now, demanding that $\frac{{\cal P}(u)}{(1-u)}$ vanishes
as we take $u\rightarrow1$ in Eq. (\ref{solu})  
gives us the condition ${\cal H} = F $. So the form of ${\cal P}(u)$ 
simplifies
to 
$${\cal P}(u)=(1-u)^2[2 F (1 +u) - \tau^{2}]$$  
The turning points of $u$ are at $u_{max}=1$ and $u_{min}=u_0$,
where
\[
u_0= \frac{\tau^{2}}{2F} - 1.
\]
The solutions are found by 
elementary integration\cite{stump,fain}
\begin{equation}
u(s)=(1-u_0)\tanh^2 \mu (s-s_{0}) + u_0
\label{soluwrf1}
\end{equation}
\begin{equation}
\varphi(s)=-\frac{\tau (s-s_{0})}{2} - {\rm Arc}{\rm tan} 
[\frac{2\mu}{\tau}{\rm tan}h \mu (s-s_{0})],
\label{soluwrf2}
\end{equation}
where $\mu^{2} = F - \tau^{2}/4$. 
The energy of the writhing family parametrized by $F,\tau$ and $L$ is 
given by
\begin{equation}
{\cal E_W} =  - FL + 8\mu \tanh \frac{\mu L}{2}-2\pi\tau {\cal 
W}_{CW}(\tau).
\label{writhing}
\end{equation}
Since we have already established
\cite{papa} 
that the 
writhing 
family are ``good curves''\cite{neukirch} 
we may compute the writhe using the simpler ``Fuller formula'' ${\cal 
W}_{F}$, which (apart from normalization)
measures the solid angle swept out by the unique shorter 
geodesic connecting the tangent vector to the north pole.  
We find by a straightforward calculation that the writhe of 
the
writhing family is
\begin{equation}
{\cal W}_{CW} = {\cal W}_{F} = \frac{2 \sgn{\tau}}{\pi} {\rm 
arc}{\rm tan}
[\frac{2\mu}{|\tau|}{\rm tanh} \frac{\mu L}{2}],
\label{writhew}
\end{equation}
in agreement with \cite{sethnadaniels}.
Including the twist energy Eq.(\ref{twistenergee}) $-\tau^2L/(2C)$, the 
``Gibbs'' free 
energy 
${\cal G}(F,\tau)$ 
of the writhing family in the mechanics approximation is 
\begin{equation}
{\cal G}_{\cal W}^{\rm Cl}(F,\tau)=-FL+8\mu\tanh{[\frac{\mu L}{2}]}-2\pi\tau 
{\cal 
W}_{CW}(F,\tau)-L\tau^2/(2C)
\label{gibbswrithe}
\end{equation}
where ${\cal W}_{CW}$ is given by Eq.(\ref{writhew}). 
Correspondingly, the ``Gibbs'' free energy
${\cal G}(F,\tau)$ 
of the straight line family in the mechanics approximation is 
\begin{equation}
{\cal G}_{SL}^{\rm Cl}(F,\tau)=-FL-\frac{L\tau^2}{2C}
\label{gibbsstraight}
\end{equation}
These two Gibbs free energies are plotted in Figure 1  as a function of 
$\tau$ for a fixed force. In our 
expressions Eq.({\ref{writhew},\ref{gibbswrithe}) for 
the energy and writhe 
of the writhing 
family, we have retained expressions like $tanh{[\mu L/2]}$, which 
become $1$ in the long polymer limit provided $\mu$ is not small. 
If one drops  these expressions our energy difference coincides with 
\cite{fain}
after correcting for an overall factor ($\sqrt{F/2}$), a presumed 
misprint in \cite{fain}.
\begin{figure}
\includegraphics{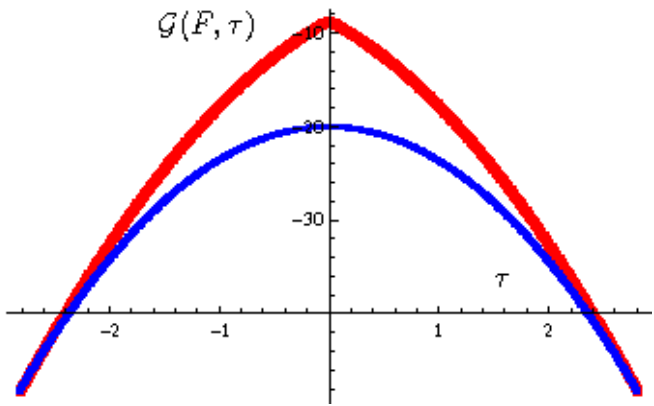}
\caption{Gibbs free energies 
${\cal G}(F,\tau)$ (from a purely mechanical approach) of the writhing 
family (thick red curve) and of the straight line family (thin blue line) 
plotted against torque $\tau$ for $L=10$, 
$F=2$ and twist elastic constant $C=1.4$. $A,k_BT,L_P$ are set to 1. 
Note that the writhing family has higher energy and that this
difference narrows as one nears buckling.}
\label{energyfig}
\end{figure}

The writhing family always has a higher
energy \cite{fain} than the straight line family for fixed force and 
torque. However, this difference is not extensive in the length
 $L$ and remains finite as $L\rightarrow\infty$.
The energy difference at zero torque is $8 \sqrt{F A}$. At higher torques
(see Fig. 1) this energy difference narrows and disappears 
at the buckling torque ($\sqrt{4FA}$, for long polymers).   
At the buckling torque, the writhing solution 
merges with the straight line and there are no classical solutions 
beyond  this torque.

The writhing family becomes relevant when there is a need to accommodate
writhe. As one can 
see by toying with a tube, applying Link to the ends of the 
tube results in deformations which cause the tube to deviate 
considerably from the 
straight line. If one applies a fixed link to the tube, the elastic 
response of the tube depends on $C$, the twist elastic constant.
In Fig. 2,3, we plot the relative extension $\xi=z/L$ versus link for a 
fixed force 
for 
large and small values of $C$. If $C$ is large (Fig. 2), there is a 
critical link above which
the SLF ceases to exist. The WF is then 
the only solution to the mechanics problem.
Further increasing the link causes the writhing family to buckle into 
plectonemes. 
Thus the writhing family represents the transition from the 
straight line to plectonemes.

\begin{figure}
\includegraphics{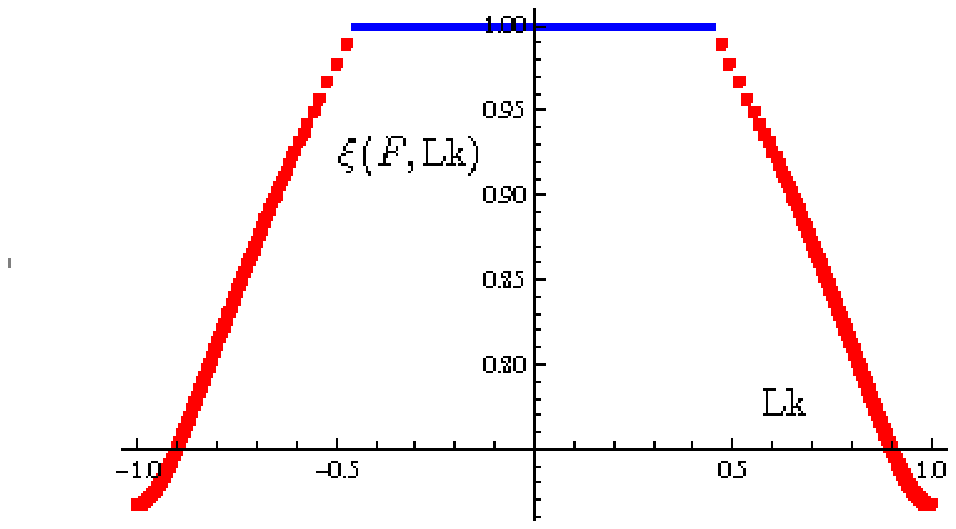}
\caption{Relative extension $\xi$ of the polymer versus applied link for
the straight line family (thin blue line) and the writhing family
(thick red line) for a stretching force of $F=2$ (in dimensionless units 
of $k_BT/L_P$) and $L=10$. 
This plot has a twist elasticity 
$C=10$ in dimensionless units. At high 
Link the straight line family ceases to exist; only the writhing
family can store writhe. 
}
\label{extlinkC10}
\end{figure}

\begin{figure}
\includegraphics{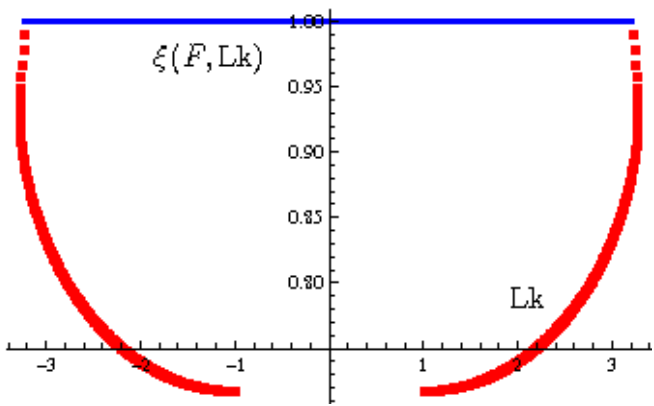}
\caption{Relative extension $\xi$ of the polymer versus applied link for 
the straight line family (thin blue line) and the writhing family 
(thick red line). This plot uses C=1.4 and a stretching force $F=2$ and 
$L=10$. 
Similar curves are seen in 
simulations \cite{neukirchmech}.} 
\label{extlinkC1}
\end{figure}

For smaller values of $C$,
the behavior is as shown in Fig.3, which plots the relative extension of 
the 
polymer versus link. 
If one takes a straight ribbon and twists it, it remains straight and
acquires link, following the thin blue line in the figure.  At a 
critical torque, the straight line meets the writhing family (thick
red line in the figure) and becomes unstable, follows the 
red line of the writhing family and buckles.
This is exactly the behavior seen in the simulations of 
Ref.\cite{neukirchmech}. See the section of the curve from $b_0$ to $b_1$
in Figure 2 of 
\cite{neukirchmech}. The rest of the curve in this 
figure of Ref.\cite{neukirchmech} is not relevant to our analysis since it 
involves self contact of the 
polymer and goes beyond our analytic approach.  A drop in extension
at buckling has also been seen in the experiments of \cite{sethnawang}.
\section{Fluctuations}
\label{fluctuations}
In a thermal environment, the polymer fluctuates around its minimum energy
configuration. This contributes a fluctuation term to the partition 
function and gives a thermal correction to the classical energy computed
in the last section.
The elastic properties of the polymer are profoundly affected by these 
fluctuations. Interpretation of the experiments on DNA elasticity 
\cite{strick} {\it requires} an inclusion of thermal effects. By
taking appropriate derivatives of the free energy computed here
one can find the theoretical predictions of the worm like chain model
for the experimentally accessible extension versus link or torque twist 
relations.
(See \cite{sethnawang} for experimental graphs of these.)

For a system with a finite number of degrees of freedom, $q^i,i=1..N$ 
with a potential $V(q)$, the mechanical energy in equilibrium is given 
by the value of the potential $V(q^*)$ at its minimum $q^*$  
 where 
\begin{eqnarray}
\partial V|_{q^*}=0\\
\label{eom}
\partial \partial V|_{q^*}>0
\label{mini}
\end{eqnarray}
where we formally write the gradient of $V$ as $\partial V$ and the 
Hessian matrix as $\partial \partial V$. (The Hessian is positive
at local minima. )
By performing a Gaussian
integral, we find that the free energy in the presence of thermal 
fluctuations is
\begin{equation}
{\cal F}=V(q^*)+\frac{1}{2} k_BT \log \det \partial \partial V|q^*.
\label{freeenergy}
\end{equation}
Our objective here is to present a calculation of the free energy of
twisted semiflexible polymers by considering fluctuations about
the mechanical solutions of the last section.
At fixed force and torque we have the minimum energy classical solution, 
which satisfies $\delta{\cal E}_{bs} = 2\pi\tau\delta{\cal W}_{CW}$, 
where ${\cal W}_{CW}$ 
is the writhe and ${\cal E}_{bs}$ is the bending and stretching energy. 
Formally, what we wish to do is to expand the 
energy and writhe around the classical configuration retaining terms to 
quadratic order. 
The quadratic integral is a functional Gaussian integral which
can then be performed and gives an answer in terms of the determinant of
the fluctuation operator.
The problem reduces to computing the determinant of the fluctuation operator. 

Computing the functional determinant of the WLC polymer involves some 
subtleties. If one 
evaluates the determinant as an infinite product of eigenvalues, one 
finds that the product does not converge and needs regularization. 
This is an artifact of the fact that we have modeled the polymer
as a space curve (or ribbon), which has an infinite number of degrees
of freedom. Physically, we know that the polymer consists of a finite
number of atoms and has finite free energy.
There is a well developed mathematical theory to deal with such 
situations. 
Multiplicative  (additive) constants can be dropped in evaluating the 
partition function (free energy) and the infinities do not affect physical 
predictions. Another subtlety arises from zero modes of the classical 
solution. If the energy functional (and boundary conditions) has 
continuous symmetries which are
broken by the classical solution, there are ``Goldstone modes'' which cost
no energy and have zero eigenvalue. Such zero modes need separate 
handling. We will face both these problems below.

Let us now consider a family of polymer configurations parametrized by 
$\sigma$. The first variation of the energy is (where $'$ denotes the 
$\sigma$ derivative). 
\begin{equation}
{\cal E}^{\prime}[\hat{t}(s)] = \int^{L}_{0} (-\ddot{\hat{t}} \cdot 
\hat{t}^{\prime} - \vec{F} \cdot \hat{t}^{\prime})\; ds
\label{fnfirstvar}
\end{equation} 
and the first variation of the writhe is
\begin{equation}
{\cal W}^{\prime}[\hat{t}(s)] = \frac{1}{2\pi} \int^{L}_{0} (\hat{t}(s) \times 
\dot{\hat{t}}(s)) \cdot {\hat{t}}^{\prime}(s)\; ds.
\label{wrifirstvar}
\end{equation} 
The Euler-Lagrange equations read ${\cal E}^{\prime} - 2\pi \tau {\cal W}^{\prime} = 0$ ~~or 
\begin{equation}
\int^{L}_{0}\{-\ddot{\hat t} - \vec{F} - \tau (\hat{t}(s) \times \dot{\hat 
t}(s))\} \cdot {\hat t}^{\prime}ds = 0 
\label{ELprime}
\end{equation} 
for arbitrary $\hat{t}^{\prime}$ satisfying 
$\hat{t}^{\prime} \cdot \hat{t} = 0$. 

\noindent
Now consider the second variation of the energy
\begin{equation}
{\cal E}^{\prime\prime} [\hat{t}(s)] = \int^{L}_{0} 
\dot{\hat t} \cdot \dot{\hat t}^{\prime\prime} - 
\int^{L}_{0} {\vec F} \cdot {\hat t}^{\prime\prime} + 
\int^{L}_{0} \dot{\hat t}^{\prime} \cdot \dot{\hat t}^{\prime}\;ds 
\label{envecvar}
\end{equation} 
and the writhe
\begin{equation}
{\cal W}^{\prime\prime}[\hat{t}(s)] = \frac{1}{2\pi} 
\int^{L}_{0} (\hat{t} \times \dot{\hat{t}}) \cdot {\hat{t}}^{\prime\prime}\;\;ds + 
\frac{1}{2\pi} \int^{L}_{0} (\hat{t}(s) \times \dot{\hat{t}}^{\prime}) 
\cdot {\hat{t}}^{\prime}\;\;ds.
\label{wrirecvar}
\end{equation} 
Computing ${\cal E}^{\prime\prime} - 2\pi\tau {\cal W}^{\prime\prime}$ we find 
that the ${\hat t}^{\prime\prime}$ 
terms combine, thanks to the Euler - Lagrange equations 
Eq.(\ref{eqnofmot}), 
to give
$\gamma\int^{L}_{0} \hat{t}^{\prime} \cdot \hat{t}^{\prime}\;ds$. (Use 
the identity $\hat{t} \cdot \hat{t}^{\prime\prime} 
+ \hat{t}^{\prime} \cdot \hat{t}^{\prime} = 
0$). 

\noindent
The form of the second variation functional is 
\begin{equation}
{\cal E}^{\prime\prime} - 2\pi\tau {\cal W}^{\prime\prime} = \int^{L}_{0} 
\{\dot{\hat{t}}^{\prime} \cdot \dot{\hat{t}}^{\prime} - \tau 
(\hat{t}_{cl}(s) 
\times 
\dot{\hat{t}}^{\prime})\cdot \hat{t}^{\prime} + \gamma_{cl}(s) 
\hat{t}^{\prime} \cdot 
\hat{t}^{\prime}\}\; ds
\label{fntotalsecvar}
\end{equation} 
This is a quadratic form in $\hat{t}^{\prime}$ which we assume is non 
negative (since we suppose that $\hat{t}_{cl}(s)$ is a {\it minimum} 
of the energy, either local or global). The subscript on 
$\hat{t}_{cl}(s)$ and $\gamma_{cl}(s)$ indicates that it is 
the classical solution about which we wish to compute the determinant of this 
quadratic form. Varying 
$\hat{t}^{\prime}$ to find the fluctuation operator (analogous to 
$\partial \partial V$ 
above) we find the eigenvalue equation (for ${\hat t}^{\prime}$ 
satisfying 
${\hat t_{cl}}.{\hat t}^{\prime}(s)=0$.) 
\begin{equation}
{\hat{\cal O}} \hat{t}^{\prime} = -\ddot{\hat t}^{\prime} - \tau \hat{t}_{cl} 
\times \dot{\hat {t}^{\prime}} + \gamma_{cl} \hat{t}^{\prime} = \lambda 
\hat{t}^{\prime} 
\label{fneveqn}
\end{equation} 
which gives us the spectrum of the fluctuation operator. If we could 
find all the eigenvalues of ${\hat {\cal O}}$, we could compute its
determinant. While this works in special cases (like the SLF) it not
usually possible in general.

To compute the determinant of the fluctuation operator $\hat {\cal O}$, 
we use a technique due to Gelfand and Yaglom (GY) \cite{gelfand,muratore}. 
We 
consider the energy Eq.(\ref{energy}) as a function of the boundary 
conditions at the end of the polymer. (This is similar to considering 
the 
Action in classical mechanics as a function of the end points in 
Hamilton Jacobi theory.)
We consider classical solutions 
which go from $\hat{t}_{i}$ at $s=0$ to $\hat{t}_{f}$ at $s=L$. Although 
our  physical problem  fixes $\hat{t}_{i}$ and  $\hat{t}_{f}$, 
the calculational technique (GY)  requires that 
we  consider variations of the energy under such changes. 
The result of GY is that   
the computation of an infinite 
dimensional functional determinant 
(the determinant of the operator 
$\hat {\cal O}$) reduces to the computation of a finite dimensional one.
The final answer is expressed in terms of the variation of the classical 
energy with respect to variations of the boundary conditions. Let ${\cal 
E}({\hat t}_i,{\hat t}_f,L)$ be the energy of the configuration of 
length $L$ that goes from ${\hat t}_i$ to ${\hat t}_f$  (where ${\hat t}_i
$ and ${\hat t}_f$ are the small variations of the boundary conditions 
about the actual boundary conditions in our problem). 
The infinite dimensional functional determinant after regularization is
reduced to the computation of a simple finite dimensional determinant:
\begin{equation}
{\rm det}{\hat {\cal O}}=\big[{\rm det} \frac{\partial^2{\cal 
E}}
{{\partial \hat t}_i \partial {\hat t}_f}\big]^{-1}
\label{2by2}
\end{equation}
evaluated at the physical boundary data.
From Hamilton-Jacobi theory we can write the quantity in 
square brackets in Eq.(\ref{2by2}) as
\begin{equation}
{\rm det}\frac{\partial p_f^\alpha}{\partial t_i^\beta}
\label{dpdt1}
\end{equation}
where $p_f^\alpha$ is the ``final momentum'' $ {\vec p}(L)$ of the 
classical configuration of length $L$ that reaches ${\hat t}_f$ from 
${\hat t}_i$. 

The method of GY is used in many fields of physics (see for 
example \cite{dunne}) for 
computing 
fluctuation determinants. Ref.\cite{emanuel} contains an application of
the technique to polymer physics in which it is used to compute the 
thermal corrections to Euler buckling  of polymers under compression. 
Our present application is to stretched and twisted polymers. The great 
advantage of the GY technique is that one does not have to diagonalize 
the fluctuation operator ${\hat {\cal O}}$ to compute its determinant. 
All one has to do is to linearise the Euler Lagrange equations 
and study the Jacobi fields. In many cases, it is not practical to 
diagonalize the fluctuation operator, while computing the Jacobi fields
is a much easier task.

We saw in the last section that the minima of the 
energy come in two families, the straight line family and the writhing 
family. 
Since the GY technique may not be familiar to many readers, we 
illustrate its use by applying it to  a case where the answer is known,
the straight line family. We calculate the fluctuation determinant by 
two methods, first explicitly diagonalizing the fluctuation operator and then by
the GY technique. 
We then apply the GY idea to the writhing 
family 
to obtain new results.

\subsection{Straight line Family} 
\label{straightline}
For the straight line family,
$\hat{t}_{cl}(s) = \hat{z}$ (for any ${F}, \tau$). 
As is well known\cite{fain,nelson}, 
there is a range of 
(${F}, \tau$) ($\tau < \tau_{c} = \sqrt{4 {F} + (\frac{\pi}{L})^{2}}$)) 
for 
which the straight 
line family is stable against perturbations. The 
eigenvalue equation Eq.(\ref{fneveqn}) reads 
\begin{equation}
- \ddot{\hat t}^{\prime} - \tau (\hat{z} \times \dot{\hat t}^{\prime}) + 
{F} \hat{t}^{\prime} = \lambda \hat{t}^{\prime},
\label{stlineeveqn}
\end{equation}
where ${\hat t}^{\prime}.{\hat z}=0$
The eigenvalues are easily worked out and the thermal correction to 
the free energy computed. This involves regularizing a divergent sum 
$2{k_{B}T}\sum_{n=1}^{\infty}{log(\mu^2+\frac{n^2\pi^2}{L^2})}$, which 
can be done by standard methods.
The final answer turns out to be ${\rm det} {\hat {\cal O}}=(\sinh{\mu 
L}/\mu)^2$ which leads to 
the thermal correction to the free 
energy 
\begin{equation}
k_BT \log{\big[\frac{\sinh{L\mu}}{\mu}\big]}
\label{stlinefluct}
\end{equation}
in agreement with Ref.\cite{abhijit}.

This computation can also be performed by line arising the
Euler Lagrange equations about the straight line. After some simple
transformations, it reduces to the inverted oscillator (in a magnetic 
field) and the solutions can be written in terms of hyperbolic
functions (unlike the trigonometric functions of the usual oscillator).
Varying the relations between final and initial data, we easily find

\begin{eqnarray}
\delta Z(L)_=&(\exp{-i\tau L/2}) \big[(\cosh{\mu L}) \delta Z(0) +  
\mu^{-1}(\sinh{\mu L}) \delta P(0)]\\
\delta P(L)_=&(\exp{-i\tau L/2}) [\mu (\sinh{\mu L}) \delta Z(0) + 
 (\cosh{\mu L}) \delta P(0)\big]
\label{stlinesolutions}
\end{eqnarray}
where $Z={\hat t}_x+i{\hat t}_y$ are complex coordinates
in the tangent plane at the north pole and $P$ the conjugate
momenta.  Setting $\delta Z(L)=0$ and solving for $\delta P(L)$,
performing the differentiation Eq.(\ref{dpdt1}) yields the determinant
${\rm det}{\hat {\cal O}}=\big[{\rm det}\frac{\partial P(L)}{\partial 
Z(0)}\big]^{-1}=(\sinh{\mu L}/\mu)^2$ as before.
Computing $k_BT/2 \log{det |\frac{\partial P(L)}{\partial 
Z(0)}|^2}$gives 
us the same answer as Eq.(\ref{stlinefluct}). 

\subsection{Writhing Family}
\label{writhingfamily}
For the writhing family, the eigenvalue equation Eq.(\ref{fneveqn}) is 
considerably more complicated (note the presence of the functions 
${\hat t}_{cl}(s)$ and $\gamma_{cl}(s)$) and the 
eigenvalues and eigenvectors are not 
available in closed form, but we can still use the GY method. The 
writhing family breaks the azimuthal symmetry (rotation about the $\hat 
z$ axis) that is present in the energy and the boundary conditions. As a 
result the solution
has zero modes which need careful handling. There are two routes open
to us and we briefly describe both of them. We can alter the boundary 
conditions so that they break the azimuthal invariance. 
For instance, we could make at least one  of ${\hat t}_i$ and ${\hat 
t}_f$ not point along the ${\hat z}$ direction.  The other route is to 
exploit 
the symmetry of the rotational invariance and treat the azimuthal 
degrees of freedom exactly. 
This leaves a one dimensional problem ($\theta$ or $u$), which we
can treat either approximately or exactly, depending on the desired
degree of accuracy.
Since the calculations are involved if 
straightforward, we outline the method and present the final answer.

\subsection{Asymmetric Boundary Conditions} 
\label{asymmetric}
We suppose that 
$u_i,\varphi_i$ 
and  $u_f,\varphi_f$, the initial and final tangent vectors are 
not {\it both} in the ${
\hat z}$ direction.  
We make a 
variation in the initial positions, while holding the final positions 
fixed. We are interested in the resultant variation in the final momenta.
To compute this, we proceed as follows exploiting the integrability
of the system.
We will keep the final configuration $(u_{f}, \varphi_{f})$ fixed and 
perform variations in $(u_{i}, \varphi_{i})$ the initial 
configuration and look at the variations in $\Phi$, $L$ and 
${\cal E}$. 
These are
\begin{eqnarray}
\delta L &=& \frac{\epsilon_i\delta u_{i}}{f(u_{i})} + \frac{\partial 
L}{\partial 
{\cal H}}\; \delta {\cal H} + \frac{\partial L}{\partial J}\; \delta 
J\nonumber\\
\delta\Phi &=& \frac{\epsilon_i\delta u_{i}}{f(u_{i})}\;h(u_{i}) + 
\frac{\partial\Phi}
{\partial{\cal H}}\; \delta {\cal H} + \frac{\partial\Phi}{\partial J}\; 
\delta 
J\nonumber\\
\delta{\cal E}&=& \epsilon_i f(u_{i}) \delta u_{i} + \delta J \Phi + 
J\delta\Phi - 
\delta{\cal H}L - 
{\cal H}\delta L.
\label{variations}
\end{eqnarray}
The Jacobian of interest is Eq.(\ref{dpdt1}), which can be expressed as 
(writing $p_{f}$ 
for $p_{u_f}$ and $J_{z}$ for $p_{\varphi_{f}}$)
\begin{eqnarray*}
\frac{\partial(p_{f}, J_z)}{\partial(u_{i}, \varphi_{i})} = 
\frac{\partial(p_{f}, J_z)}{\partial({\cal H}, J_z)}\;\;
\frac{\partial({\cal H}, J_z)}{\partial(u_{i}, \varphi_{i})}.
\end{eqnarray*}
From 
\begin{eqnarray*}
\epsilon_f \delta p_{f} = \delta f(u_{f})/(1-u^{2}_{f}) = \frac{\delta 
{\cal 
H}}{f(u_{f})} - \frac{h(u_{f})}{f(u_{f})}\;\;\delta J_z
\end{eqnarray*}
we find
\begin{equation}
\left(\begin{array}{c}\delta p_{f}\\ 
                              \delta J \end{array}\right) =  
\left(\begin{array}{cc}\frac{\epsilon_f}{f(u_{f})} & 
\frac{-h(u_{f})\epsilon_f}{f(u_{f})}\\ 
0 & 1\end{array}\right)\;\;\left(\begin{array}{c}\delta {\cal H}\\ 
                              \delta J_z \end{array}\right) = {\cal 
A} 
\left(\begin{array}{c}\delta {\cal H}\\ 
                              \delta J_z \end{array}\right)
\label{pJtoEJ}
\end{equation}
where ${\cal A}$ is a $2 \times 2$ matrix defined by Eq.(\ref{pJtoEJ}).\\

\noindent
Setting 
\(\delta L = 0\) and \(\delta \Phi = \delta(\varphi_{2} - \varphi_{i}) 
= -\delta\varphi_{i}\) in Eq.(\ref{variations}) we find
\begin{equation}
{\cal D} \left(\begin{array}{c}\delta {\cal H}\\ 
                              \delta J \end{array}\right) =  
\left(\begin{array}{cc}\frac{\partial L}{\partial {\cal H}} & \frac{\partial L}{\partial J}\\ \frac{\partial \Phi}{\partial {\cal H}} & 
\frac{\partial \Phi}{\partial J}\end{array}\right)\;\;\left(\begin{array}{c}\delta {\cal H}\\ 
                              \delta J \end{array}\right)
\end{equation}
\begin{equation}
= \left(\begin{array}{cc}\frac{-\epsilon_i}{f(u_{i})} & 0\\ 
                         \frac{-h(u_{i})\epsilon_i}{f(u_{i})} & 
-1\end{array}\right)   
\left(\begin{array}{c}\delta u_{i} \\
                      \delta \varphi_{i}\end{array}\right) = {\cal C}  
\left(\begin{array}{cc}\delta u_{i} \\
                       \delta \varphi_{i}\end{array}\right)
\label{chain}
\end{equation}
where the $2 \times 2$ matrices ${\cal D}$ and ${\cal C}$ are defined in
Eq. (\ref{chain}). From 
\begin{equation}
({\rm det} {\hat{\cal O}}) ^{-1} = 
{\rm det} \left(\frac{\partial p_{f}}{\partial q_i}\right),
\end{equation}
we finally arrive at 
\begin{eqnarray}
{\rm det} {\hat{\cal O}} &=& (\det {\cal A})^{-1} (\det {\cal D}) (\det {\cal C})^{-1}\nonumber\\
       &=& f(u_{i}) f(u_{f}) \det {\cal D} \\
       &=&f(u_{i}) f(u_{f})\big[\frac{\partial L}{\partial \cal{H}}\frac{\partial \Phi}{\partial 
J}-\frac{\partial \Phi}{\partial \cal{H}}\frac{\partial L}{\partial 
J}\big]\epsilon_i\epsilon_f
\label{determ}
\end{eqnarray}
where ${\cal D}$ is the $2 \times 2$ matrix defined in Eq.(\ref{chain}). The final
answer for the total free energy including thermal corrections is
\begin{equation}
{\cal G}(F,\tau,k_BT)={\cal E}_{cl}+\frac{k_BT}{2} 
\log{\big[\epsilon_i\epsilon_ff(u_i)f(u_f) 
\big[\frac{\partial L}{\partial \cal{H}}\frac{\partial \Phi}{\partial 
J}-\frac{\partial \Phi}{\partial \cal{H}}\frac{\partial L}{\partial 
J}\big]\big]}\
\label{asymthermalcorr}
\end{equation}
These derivatives are evaluated at constant $u_i,u_f$. 
While this final answer may appear to be abstract, it is in fact
quite amenable to numerical methods.
All the functions appearing here 
can be expressed 
Eq.(\ref{action},\ref{length},\ref{phiaction}) in 
terms of 
elliptic integrals and numerically 
evaluated.

Although Eq.(\ref{asymthermalcorr}) is written for asymmetric boundary 
conditions, the thermal correction to the free energy for long polymers 
with symmetric boundary conditions can be teased out of it, since the 
free energy for long polymers is expected to be independent of the 
boundary conditions. We take the limit of $u_i\rightarrow 1$, which
leads to $f(u_i)$ vanishing. In Eq.(\ref{asymthermalcorr}) the terms that
multiply $f(u_i)$, we 
need only 
keep those terms that diverge as $u_i\rightarrow 1$. This 
results in a simple
expression for the leading correction to the free energy:
\begin{equation}
\Delta {\cal G}_W = k_BT L \mu=k_BT L \sqrt{F-\tau^2/4}
\label{thermalwrithe}
\end{equation}
In Eq.(\ref{thermalwrithe}) we have dropped lower order terms which 
are not extensive in $L$. Notice that the numerical value of the
thermal correction to the free energy is the {\it same} as  for the 
straight line family. This can be understood physically since 
for any finite $\mu$, most of the polymer is straight. 
We would not in general expect the fluctuation energies of SLF and WF to 
be equal in short polymers.
The expressions Eqs. (\ref{determ}-\ref{thermalwrithe}) contain the 
main results of this study; an analytic calculation of the thermal
corrections to the free energy.

Our main result expressing the fluctuation determinant in simple terms
sheds considerable light on the stability of polymer configurations.
Note that the determinant vanishes as either $u_i$ or $u_f$ approaches a
turning point. This signals the appearance of null eigenvalues for the
Hessian or Fluctuation operator Eq.(\ref{fneveqn}). Such effects have 
been well studied in optics and mechanics \cite{arnold}, where they
reveal the appearance of caustics, focusing  and conjugate points. If  
the polymer configuration goes from $u_i$ to $u_f$ passing through turning
points, it is not stable against small perturbations. The Gibbs free 
energy ${\cal G}(\tau,F)$ can be lowered at constant torque by such 
perturbations. At constant torque, such configurations will decay to 
lower energy configurations.

However, if one works in the corresponding Helmholtz ensemble, at 
constant link, which is how most experiments
are done, the allowed perturbations are  
also required to maintain constant writhe (supposing for simplicity, 
$C$ to be infinite) and this instability
disappears. However, if $u_i$ and $u_f$ are separated by {\it two}
turning points,  there are {\it two} independent perturbations,
which lower the Energy. Working to second order in the perturbation,
one of these lowers the writhe and the other increases it.  
One can take a linear combination of these two perturbations
to maintain constant writhe. Such configurations are unstable at both
constant torque and constant Link. Note that polymer configurations 
that extend for more than a period $P$ are therefore unstable to 
perturbations in accord with the findings of \cite{papa}. 
They are not local minima of either the Helmholtz or
Gibbs energies.
\section{Towards Statistical Mechanics}
Till this point we have relied on the mechanical approach, 
dealing with individual polymer configurations and considering small
fluctuations around them. This has the advantage that we know which
individual configuration we are dealing with and can show \cite{papa}
that the dominant configurations admit a local writhe formula.
We now move towards a more statistical mechanical approach  
and integrate over configurations. In the process, we lose touch
with individual configurations, but we gain by getting 
a fuller treatment. From the mechanical approach, we have gained 
confidence that the sum over configurations is dominated by ``good
curves'' that admit a local writhe formula. Using this idea, 
we will now extrapolate to more general situations where the 
thermal fluctuations are appreciable and continue to use a local
writhe formula. There 
is a logical jump involved here and we have argued elsewhere 
\cite{samsupabhi} that this leads
to a good approximation provided that the polymer is under stretch.

\subsection{Symmetric Boundary Conditions}
\label{symmetric2} 
If we assume that the boundary
conditions are azimuthally symmetric, we can exploit the symmetry
to treat the $\varphi$ degrees of freedom exactly by ``integrating them 
out''. We reduce the problem
to a single variable $\theta$ (or $u$). In the process of reduction
the effective potential picks up an extra contribution from the 
thermal fluctuations of the $\varphi$ degrees of freedom. As a result
the effective potential explicitly depends on the temperature. 
The origin of this effect is that path integration in general 
coordinates has subtleties coming from the measure\footnote{This 
can also be seen by operator methods, where it emerges from the need 
to make the reduced differential operator self adjoint.}, as 
pointed 
out by 
Edwards and Gulyaev \cite{edward}. 
These techniques have been applied \cite{rajaramanweinberg} to compute 
the quantum corrections to soliton
energies. Our calculation is mathematically similar, but physically
different since we deal with thermal corrections rather than quantum 
ones. Our treatment below closely follows these earlier studies, with 
appropriate modifications.

After integrating out the $\varphi$ degrees of freedom, the $\theta$ 
motion is governed by an effective potential 
\begin{equation}
V_{eff}(\theta)=F\cos{\theta}-\frac{k_BT}{8L_P}+\tau^2
\frac{(1-\cos{\theta})}{1+\cos{\theta}}-\frac{k_BT}{8L_P\sin^2{
\theta}}
\label{effectivepotential}
\end{equation}
and the earlier equation
Eq. (\ref{solu}) is replaced by
\begin{equation}
{\dot u}^2={\cal P}_T(u)=
(f_T(u))^2=2(({\cal H}+\frac{k_BT}{8 L_P})-F 
u)(1-u^2)-\tau^2(1-u)^2 
+\frac{k_BT}{4L_P}
\label{thermalpoly}
\end{equation}
Note the explicit temperature dependence in Eq.(\ref{effectivepotential}). 
Note that $u=1$ is no longer an allowed solution. The straight line 
is destabilized by entropic effects. Indeed it would be hard to 
distinguish between SLF and WF here because we are no longer dealing
with single configurations but ensembles of them. 
However, we can still continue to use a local writhe formula 
for torques large enough
that the turning point $c$ is positive.
Such curves have tangent vectors entirely in the northern hemisphere
and are ``good" curves.

We now treat the $\theta$ (or $u$) motion approximately by considering
the ``configurations'' $u(s)$ which satisfy the Euler-Lagrange
equations.
One finds the ``classical'' contribution by evaluating the energy 
on the ``classical'' solution. (``Classical'' is in quotes because it is 
a bit of a misnomer, since the effective potential includes 
contributions from thermal fluctuations.)
The thermal correction to the free energy can now be computed:
\begin{equation}
{\cal F}(T,F,\tau)={\cal E}_{''{\rm Cl}''}(T,F,\tau)+k_BT/2
\log{\big[-4({\cal H}-F)(\frac{\partial L}{\partial{\cal H} 
})\big]}
\label{thermalcorrectionsymm}
\end{equation}
In the first term in Eq.(\ref{thermalcorrectionsymm})
${\cal E}_{\rm ''Cl''}$ is computed using the thermally corrected
${\cal P}_T(u)$. In the second term which is already of order $k_BT$,
we have used the old ${\cal P}(u)$ since the difference is higher
order in $k_BT$. $L$ is given in terms of ${\cal H} $ by formula 
Eq.(\ref{length}).
These formulae are numerically  
tractable (analytically they are still elliptic integrals) and can be 
used to work out the predictions of the model.

\begin{figure}
\includegraphics{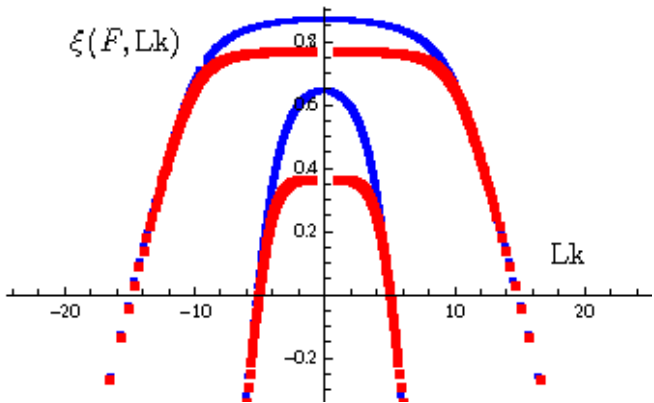}
\caption{Relative extension $\xi$ versus link for a fluctuating polymer
for the straight line family (thin blue line) and the writhing family
(thick red line). The inner curves correspond to $F=2$ and the outer
ones to $F=15$.
}
\label{hatcurves}
\end{figure}

\subsection{Statistical Approach}
\label{statistical} 
In fact, it is possible to also treat the
$u$ degree of freedom exactly but with more work. The effective 
potential Eq.(\ref{effectivepotential}) gives rise to a Schr\"odinger 
equation 
\begin{equation}
-\frac{1}{2}\frac{d^2\psi}{d\theta^2}+V_{eff}(\theta)\psi=\epsilon \psi. 
\label{schrodinger}
\end{equation}
Solving the Schr\"odinger equation gives the full
propagator, which in turn gives us the free energy of the polymer.
It is convenient to pass to imaginary ``time" 
which changes the relative sign between the kinetic and potential 
terms. This turns the potential over
$V_{eff}\rightarrow-V_{eff}$. For $\tau^2\ge \frac{k_BT}{32}$ the 
potential has an
attractive well at $\theta=\pi$ coming from the last two terms in the 
effective potential (Eq. \ref{effectivepotential}). The potential near 
$\theta=\pi$ diverges 
as 
$-1/(\pi-\theta)^2$. As a result, writhe can be stored at arbitarily 
small energy cost. 
This pathology of the model is caused by paths
that wind round the south pole and has to be regulated by imposing
a cutoff\cite{Bouchiat,fuller1}.
This is a reflection in this model of the 
plectonemic transition\cite{Bouchiat}. As these authors point out,
a cutoff corresponding to the effective diameter of the DNA molecule is
necessary.
In fact the twist experiments
reported in \cite{sethna} show the polymer making transitions between
the plectonemic phase and the normal phase. It would be interesting
to compare the predictions of this theoretical model with the 
experiment. Possible points of contact are the relative time spent
in each phase and the mean rate of transitions, both of which can 
be calculated from the model. Of course a comparison with the real 
experiment will also involve the dynamics of the bead as done in
\cite{sethnadaniels}. Our study deals purely with the polymer.

For long polymers, we can use the 
eigenstate expansion with states $\psi_n$ and energies
$\epsilon_n$
\begin{equation}
K({\hat t}_i,{\hat t}_f,L)=\Sigma_n 
\psi_n({\hat t}_i))\psi_n({\hat t}_f) \exp{\big[- \epsilon_n \big]}
\label{propagator}
\end{equation}
is dominated by the ground state and compute the free energy 
as the ground state energy $\epsilon_0$ of the Schr\"odinger operator
in Eq.(\ref{schrodinger}).

\section{Conclusion}
We have given a detailed analysis of the mechanics and fluctuations 
of writhing polymers under stretch. Our main results are 1) an explicit 
characterization of the mechanics of the writhing family, its energy and 
writhe 2) a study of fluctuations about these equilibrium 
configurations and formulae which can be used to calculate the thermal 
correction to the free energy.   
In particular, we have presented explicit approximate analytic expressions 
for the free energy 
(Eqs.(\ref{asymthermalcorr}) and \ref{thermalcorrectionsymm})
of a semiflexible polymer ribbon.
We do not have room here for a detailed
elucidation of the ``phase diagram'' which results from the 
free energy expressions above. 
However, our work shows that approximate analytical treatments of the 
problem of writhing DNA
are possible. 
Such analytical treatments are complementary to simulations
(numerical experiments) and real experiments.

A curious feature of the energy (and free energy) of the writhing family is the
cusp at $\tau=0$. On differentiation this results in a discontinuity in the link
between positive and negative values of $\tau$. The discontinuity is visible
in Figure 2 (the gap in the thick red curve) and is also present in simulations
(see Figure 2 of \cite{neukirchmech}). This feature can be traced to the fact that 
the $\tau=0$ member of the writhing family has self intersections and there is a consequent
jump in the writhe across it. 

We started this paper with a mechanical approach and moved progressively
to a more statistical mechanical point of view. The advantage of mechanics
is that we are dealing with single configurations of the polymer and
can therefore check whether the local writhe formula applies. 
The work of \cite{papa} shows that it does apply 
to all the local minima of energy (at least for a class of solutions 
($J_z=-\tau$), which includes symmetric initial data).
We expect from here that
small fluctuations about these  ``good curves'' are also ``good 
curves''. Further extrapolating this idea, we expect that the present models
based on a local writhe formula are a good approximation to the exact, but 
difficult to solve models based on the non local writhe. 
This result supports the idea 
proposed in \cite{samsupabhi} that in a statistical sense, one can 
approximate self avoiding models by south avoiding ones. 

We mentioned before that in mechanics, the straight line is incapable of accommodating writhe. 
This is shown in Figs 2 and 3, where the extension of the thin 
blue line is maximal 
independent of the applied link. This situation changes drastically when thermal fluctuations
are incorporated as in Fig. 4. The straight line family no longer has maximal extension (the
thin blue line curves down). Thus the applied link is stored 
entirely in the thermal fluctuations of
the  molecule. The molecule has helical 
fluctuations and the molecule stores writhe by  preferentially
writhing with one helicity. This is in 
sharp contrast to the writhing solutions,
which can store writhe even without thermal fluctuations.

A mechanical treatment is expected to work well for short polymers.
For long polymers without applied force, the system is entropy dominated
and therefore our treatment does not apply. For long polymers under 
a very high stretching force (for example the regime $F\approx 10$ in 
\cite{sethnawang} corresponding to $1{\rm pN}$), the entropic 
fluctuations 
are tamed and a 
mechanical approach is again possible. 
As the force is lowered, entropic effects begin to gain importance. 
Our treatment considers fluctuations 
about non perturbative solutions of the writhing family and therefore 
goes beyond perturbation theory about the straight line. 
We have argued elsewhere\cite{samsupabhi} that the 
free energy predicted by ``south'' avoiding models 
are a good approximation to that of ``self avoiding models'' provided 
the polymer is under stretch. This 
is certainly true at low link, since perturbation theory about the
straight line applies. At extremely high links, the same is true for
a different reason: the configurations in which the polymer winds 
around itself in the self avoiding model are mimicked by south
winding configurations in the south avoiding model. At intermediate
links, we expect that the polymer is in a 
coexistence phase \cite{marko} 
with some
of its length storing the writhe in the  plectonemic phase and the 
rest of it at a torque below the buckling torque. 
Our study suggests that even in this regime which is 
far from perturbative, the two models give approximately similar 
predictions. As mentioned earlier, this is based on 
our demonstration \cite{papa} that 
the dominant configurations
(the local minima of the energy) are good curves.
Only a simulation will convincingly demonstrate how low one can go 
in stretching force before the approximation breaks down.
We hope to compare these theoretical 
predictions with computer simulations in the future.\\

%\subsection{Subsection title}
%\label{sec:2}
%as required. Don't forget to give each section
%and subsection a unique label (see Sect.~\ref{sec:1}).
%\paragraph{Paragraph headings} Use paragraph headings as needed.
%\begin{equation}
%\end{equation}

%\begin{figure}
%  \includegraphics{example.eps}
%\end{figure}
%\begin{figure*}
%  \includegraphics[width=0.75\textwidth]{example.eps}
%\end{figure*}
%\begin{table}
%\begin{tabular}{lll}
%\hline\noalign{\smallskip}
%\end{tabular}
%\end{table}

\begin{acknowledgements}
It is a pleasure to thank  Michael Berry, Abhishek Dhar, Sreedhar B.
Dutta, Abhijit Ghosh, Yashodhan Hatwalne, Jaya Kumar, Viswambhar Pati,
Abhishodh Prakash, Arthur La Porta and Sanjib Sabhapandit for 
discussions.
\end{acknowledgements}

%BibTeX users please use one of
%\bibliography{referencespolymerold.bib}   % name your BibTeX data 
%base

%\begin{thebibliography}{}
% etc
%\end{thebibliography}

\end{document}